\documentclass[a4paper,fleqn,11pt]{article}
\usepackage{epsfig}
\usepackage{fancybox}

   \newcommand{\be}{\begin{equation}}
   \newcommand{\ee}{\end{equation}}

   \newcommand{\bea}{\begin{eqnarray}}
   \newcommand{\eea}{\end{eqnarray}}

  \newcommand{\bm}[1]{\mbox{\boldmath$#1$}}

\begin{document}
\begin{center}
{\bf \Large Quark Spectrum near Chiral Transition Points}
\vskip 2em
{Masakiyo Kitazawa${}^a$, 
Teiji Kunihiro${}^b$ and 
Yukio Nemoto${}^c$
\footnote{Talk given by Y. N.}
\vskip 1.5em
{\it ${}^a$Department of Physics, Kyoto University, Kyoto, 606-8502 Japan}\\
{\it ${}^b$Yukawa Institute for Theoretical Physics, Kyoto University, Kyoto
606-8502 Japan}\\
{\it ${}^c$Department of Physics, Nagoya University, Nagoya, 464-8602 Japan}}
\end{center}

\begin{abstract}
Near the critical temperature of the chiral phase transition, 
a collective excitation due to fluctuation of the chiral order parameter 
appears. 
We investigate how it affects the quark spectrum near but above the critical
temperature. 
The calculated spectral function has many peaks. 
We show this behavior can be understood in terms of resonance scatterings 
of a quark off the collective mode. 

\end{abstract}

\section{Introduction}

Quark gluon plasma (QGP) near chiral phase transition at high
temperature ($T$) and low density is recently much studied
experimentally and theoretically.
Characteristic features of QGP in that region are considered to come
from the strong coupling nature of QCD.
For example, in recent lattice QCD analysis, the lowest charmonium state
survives above the critical temperature $T_C$
\cite{Asakawa:2003re}, which indicates there exist
some hadronic bound states even in the QGP phase.
Phenomenologically the existence of hadronic states above 
$T_C$ was
suggested for the light quark sector many years ago.
If the order of the phase transition is the second or the
nearly second, there are long range correlations leading to
fluctuations.
It is shown that the fluctuation of the chiral order
parameter, $\langle \bar{\psi} \psi \rangle$, survives up to
$\varepsilon = (T_C-T)/T_C \sim 0.2$ 
at zero density\cite{Hatsuda:1984jm}.
This value is much larger than that in superconductors in metal,
$\varepsilon\sim O(10^{-3})$, which is due to the strong coupling nature
between quarks.
Inspired by the RHIC data and the lattice QCD results,
such hadronic states for both the light and heavy quarks at 
QGP near $T_c$ are recently elaborated\cite{Shuryak:2004cy}.

In this paper we investigate the quark spectrum near $T_C$ in the QGP phase.
Owing to the strong couping nature, the quark spectrum could
be much different from the free quark one.
In such an analysis, however, perturbative QCD based on the hard thermal 
loop (HTL) resummation would not be valid because $T_C$ is not so high.
Instead, we employ a low energy effective model of QCD and
incorporate physical phenomena which are important near $T_C$.
As stated above, one of important degrees of freedom near $T_c$
is the fluctuation mode of the order parameter, because it
reflects the long range correlation due to the second order
phase transition.
In the chiral phase transition, we can observe the 
fluctuation of the chiral condensate, which we call the soft mode,
near but above $T_C$.
Therefore, we investigate how the fluctuation of
the chiral condensate 
contributes to the quark self-energy and affects the spectrum
near but above $T_C$.

In Sec.\ref{sec:form}, we formulate the single-quark spectral
function which incorporates the fluctuation effects.
Numerical results and the discussions are given in 
Sec.\ref{sec:res}.
The conclusion and the outlook are in Sec.\ref{sec:conc}.

\section{Formulation} \label{sec:form}

We employ an effective model of QCD, Nambu--Jona-Lasinio model,
\be
  \mathcal{L}=\bar{\psi} i \partial \hspace{-0.5em} / \psi
  + G_S [(\bar{\psi} \psi)^2 + (\bar{\psi}i\gamma_5\vec{\tau}\psi)^2],
\ee
with $\vec{\tau}$ being the flavor SU(2) Pauli matrix.
We consider the chiral limit to investigate the second order chiral
phase transition at zero density.
The coupling constant $G_S=5.5$ GeV${}^{-2}$ and the three dimensional
cutoff $\Lambda=631$ MeV are taken from Ref.\cite{Hatsuda:1994pi}.
The critical temperature $T_C$ is determined from
the thermodynamic potential in the mean field approximation.
The critical temperature at zero density is $T_C=193.5$ MeV and
the tricritical point (TCP), where the order of the transition
changes from the second to the first, is located at $T_{TCP}=84.2$ MeV
and $\mu_{TCP}=278.6$ MeV.

Firstly we formulate the fluctuation of the chiral order parameter.
The collective excitation due to the fluctuations of
the chiral condensate above $T_C$ can be expressed by quark-antiquark
effective propagators in the scalar isoscalar ($\sigma$) 
and the pseudoscalar isovector ($\pi$) channels\cite{Hatsuda:1984jm}.
We evaluate them in the random phase approximation to give
\be
  \mathcal{D}_\sigma(\bm{p},\nu_n)=
  \frac{2G_S}{1+2G_S \mathcal{Q}(\bm{p},\nu_n)},
  \mathcal{D}_\pi(\bm{p},\nu_n)=3 \mathcal{D}_\sigma(\bm{p},\nu_n),
\ee
in the imaginary time formalism.
Here $\nu_n=2\pi n T$ is the Matsubara frequency for bosons and 
$\mathcal{Q}(\bm{p},\nu_n)$ is the one-loop quark-antiquark 
polarization function,
\be
  \mathcal{Q}(\bm{p},\nu_n)=T\sum_m \int \frac{d^3 q}{(2\pi)^3}
  {\rm Tr} [ \mathcal{G}_0(\bm{q},\omega_m)
  \mathcal{G}_0(\bm{p}+\bm{q},\nu_n+\omega_m) ],
\ee
where the free Matsubara quark propagator $\mathcal{G}_0(\bm{p},\omega_n)$
is given by
$\mathcal{G}_0(\bm{p},\omega_n)=\{(i\omega_n+\mu)\gamma_0-\bm{p}\cdot
  \bm{\gamma}\}^{-1}$
with $\omega_n=(2n+1)\pi T$ being the Matsubara function for fermions
and the trace is taken over color, flavor and Dirac indices.

\begin{figure}[t]
\begin{center}
\epsfig{file=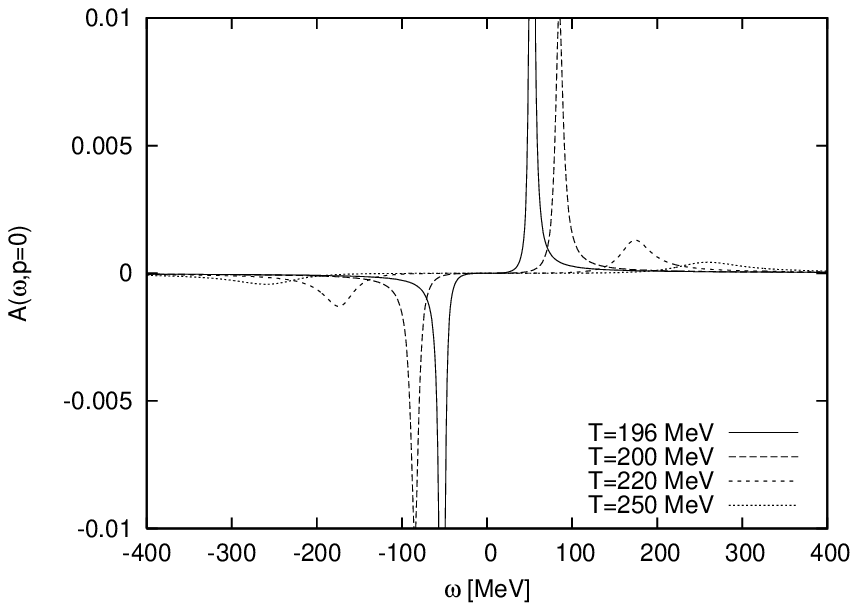, width=190pt}
\epsfig{file=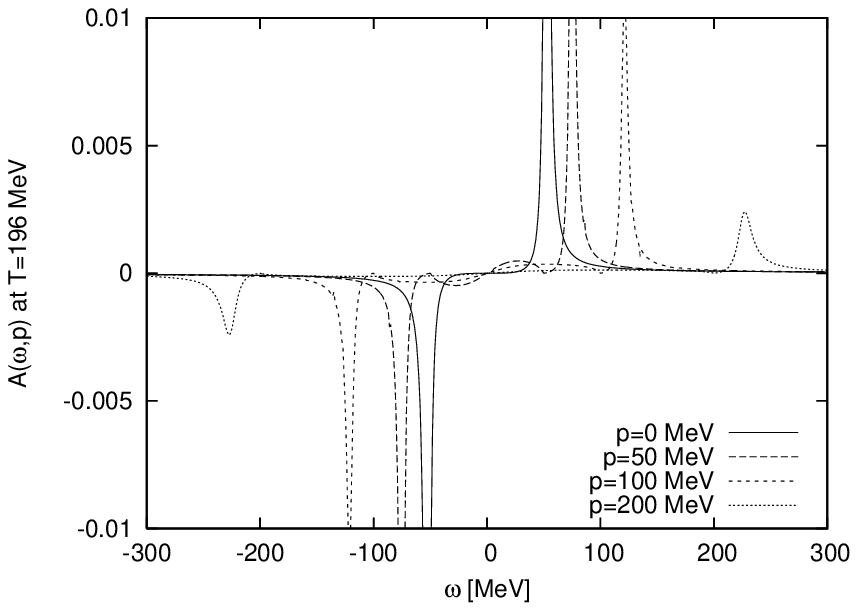, width=190pt}
\caption{Temperature dependence at $\bm{p}=0$ (left) and
momentum dependence at $T=196$ MeV($1.01T_C$)
 (right) of the spectral functions,
eq.(\ref{rho}).}
\label{soft-tp}
\end{center}
\end{figure}

After carrying out the Matsubara summation, one takes the analytic
continuation to obtain the retarded functions,
$D_{\sigma,\pi}^R(\bm{p},\omega)=
\mathcal{D}_{\sigma,\pi}(\bm{p},\nu_n)|_{i\nu_n=\omega+i\eta}$
and
$Q^R(\bm{p},\omega)=\mathcal{Q}(\bm{p},\nu_n)|_{i\nu_n=\omega+i\eta}$.
From the effective propagator $D_{\sigma,\pi}^R$, 
one can obtain the spectral
function which shows the peak due to the collective mode\cite{Hatsuda:1984jm},
\be
  \mathcal{A}_{\sigma}(\bm{p},\omega)=-\frac{1}{\pi}{\rm Im} 
  D_{\sigma}^R(\bm{p},\omega).
  \label{rho}
\ee
$\mathcal{A}_{\pi}$ is the same as the $\mathcal{A}_{\sigma}$
up to a constant factor.
We show the temperature dependence of the $\mathcal{A}_\sigma(\bm{p},\omega)$
at $\bm{p}=0$ in Fig.\ref{soft-tp}(left).
The peaks denote the collective mode which moves to the origin as the
temperature approaches $T_C$ from above.
One sees that the peaks appear up to 
$\varepsilon\equiv(T-T_C)/T_C=0.2$.
The momentum dependence of the $\mathcal{A}_\sigma(\bm{p},\omega)$
is shown in Fig.\ref{soft-tp}(right).
As the momentum lowers, the peak approaches the origin and
diverges at $T_C$ and $(\bm{p},\omega)=(\bm{0},0)$, which means
the softening.

\begin{figure}
\begin{center}
\epsfig{file=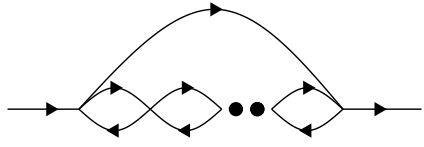, width=160pt}
\caption{The self-energy $\tilde{\Sigma}$ in the T-matrix
approximation.}
\label{sig}
\end{center}
\end{figure}

Using the quark-antiquark effective propagators formulated above, 
we incorporate the fluctuation of the chiral
condensate into the spectral function of quarks expressed as
\be
  \mathcal{A}_q(\bm{p},\omega)=\rho_0(\bm{p},\omega)\gamma^0-
  \rho_V(\bm{p},\omega)\hat{\bm{p}}\cdot\bm{\gamma},
\ee
with
$
  \rho_0(\bm{p},\omega)=\rho_-(\bm{p},\omega)+\rho_+(\bm{p},\omega),\ 
  \rho_V(\bm{p},\omega)=\rho_-(\bm{p},\omega)-\rho_+(\bm{p},\omega),
  \label{rho0v}
$
and
\be
  \rho_\mp(\bm{p},\omega)=-\frac{1}{2\pi}
  \frac{{\rm Im}\Sigma_\mp(\bm{p},\omega)}
  {[\omega\mp|\bm{p}|-{\rm Re}\Sigma_\mp(\bm{p},\omega)]^2
   +[{\rm Im}\Sigma_\mp(\bm{p},\omega)]^2}.
\ee
One notes that there is no Lorentz-scalar term in 
$\mathcal{A}_q(\bm{p},\omega)$ in the chiral limit.
$\Sigma_\mp(\bm{p},\omega)$ are obtained from the 
retarded quark self-energy,
\be
  \Sigma^R(\bm{p},\omega)
  = \gamma^0(\Sigma_- \Lambda_- + \Sigma_+ \Lambda_+),
\ee
with $\Lambda_{\mp}=(1\pm\gamma^0\hat{\bm{p}}\cdot\bm{\gamma})/2$.
Effects of the fluctuation on the quark spectrum are incorporated
through the quark self-energy for which
we employ the non-self-consistent T-matrix approximation
\cite{Kadanoff, Kitazawa:2005vr},
\be
  \tilde{\Sigma}(\bm{p},\omega_n) =
  T\sum_{m}\int\frac{d^3 q}{(2\pi)^3} 
  \mathcal{D}(\bm{p}-\bm{q},\omega_n-\omega_{m_1}) 
  \mathcal{G}_0(\bm{q},\omega_{m})
\ee
with
$\mathcal{D}(\bm{p},\omega_n)=\mathcal{D}_\sigma(\bm{p},\omega_n)
+\mathcal{D}_\pi(\bm{p},\omega_n)$.
Fig.\ref{sig} is the diagrammatic expression for the $\tilde{\Sigma}$.
After the summation of the Matsubara frequency and the
analytic continuation, $i\omega_n\to \omega+i\eta$,
it becomes the retarded function,
\bea
  \lefteqn{\Sigma^R(\bm{p},\omega)}\nonumber \\ &=& 
  \int\frac{d^4q}{(2\pi)^4} \bigg[
  -\coth\left(\frac{q_0}{2T}\right){\rm Im}
  D^R(\bm{p}-\bm{q},-q_0)
  G_0^R(\bm{q},q_0+\omega+i\eta) \nonumber \\
  & & +\tanh\left(\frac{q_0}{2T}\right)
  D^R(\bm{p}-\bm{q},\omega-q_0+i\eta) 
  {\rm Im}G_0^R(\bm{q},q_0) \bigg] \nonumber \\
  &=& -\frac{1}{2}\int\frac{d^4q}{(2\pi)^4}
  \frac{{\rm Im}D^R(\bm{p}-\bm{q},q_0)}{q_0-\omega+|\bm{q}|-\mu-i\eta}
  (\gamma^0-\hat{\bm{q}}\cdot\bm{\gamma})
  \left[ \coth\left(\frac{q_0}{2T}\right) 
  + \tanh\left(\frac{|\bm{q}|-\mu}{2T}\right) \right]
  \nonumber \\
  & & +\frac{1}{2}\int\frac{d^4q}{(2\pi)^4}
  \frac{{\rm Im}D^R(\bm{p}-\bm{q},q_0)}
  {q_0-\omega-|\bm{q}|-\mu-i\eta}
  (-\gamma^0-\hat{\bm{q}}\cdot\bm{\gamma})
  \left[ \coth\left(\frac{q_0}{2T}\right) 
  + \tanh\left(\frac{-|\bm{q}|-\mu}{2T}\right) \right], \nonumber \\
  \label{self-ene1}
\eea
with $\hat{\bm{q}}=\bm{q}/|\bm{q}|$.
The retarded Green function for a massless quark is expressed as
\be
  G^R(\bm{p},\omega)=
  \frac{\Lambda_-\gamma^0}{\omega-|\bm{p}|-\Sigma_-+\mu+i\eta}+
  \frac{\Lambda_+\gamma^0}{\omega+|\bm{p}|-\Sigma_++\mu+i\eta}.
  \label{g-r1}
\ee
It is clear that for a free particle $(\Sigma_\mp=0)$,
the poles in each term in the RHS of eq.(\ref{g-r1}) give the
dispersion relations for a free quark and a free antiquark, respectively.
At finite temperature and density, there can be several poles in each
term.

\section{Results and Discussions} \label{sec:res}

\begin{figure}
\begin{center}
\epsfig{file=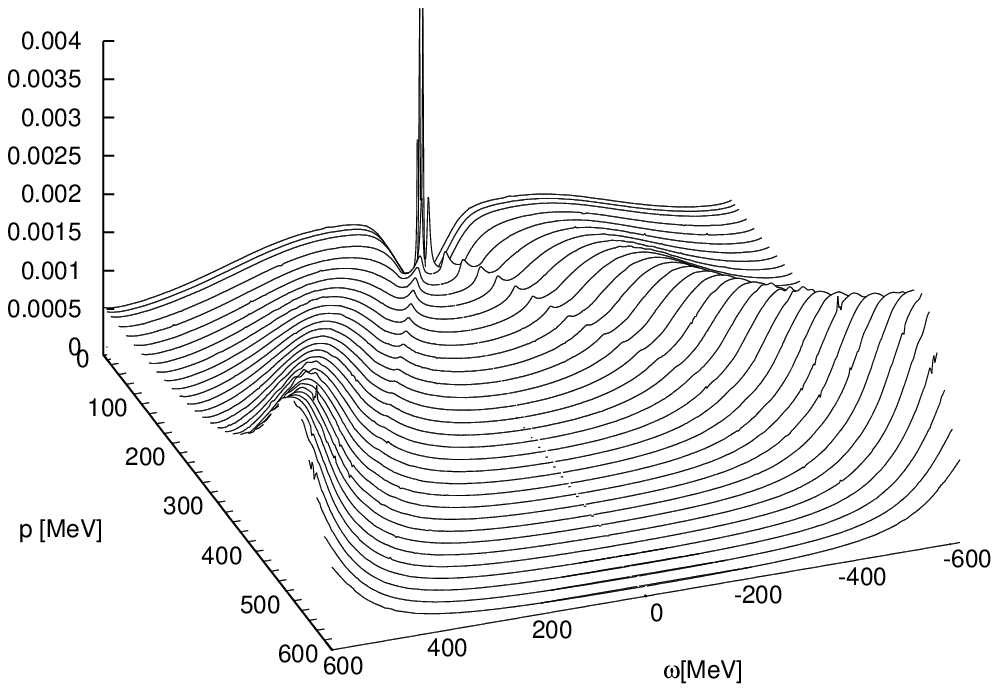, width=190pt}
\epsfig{file=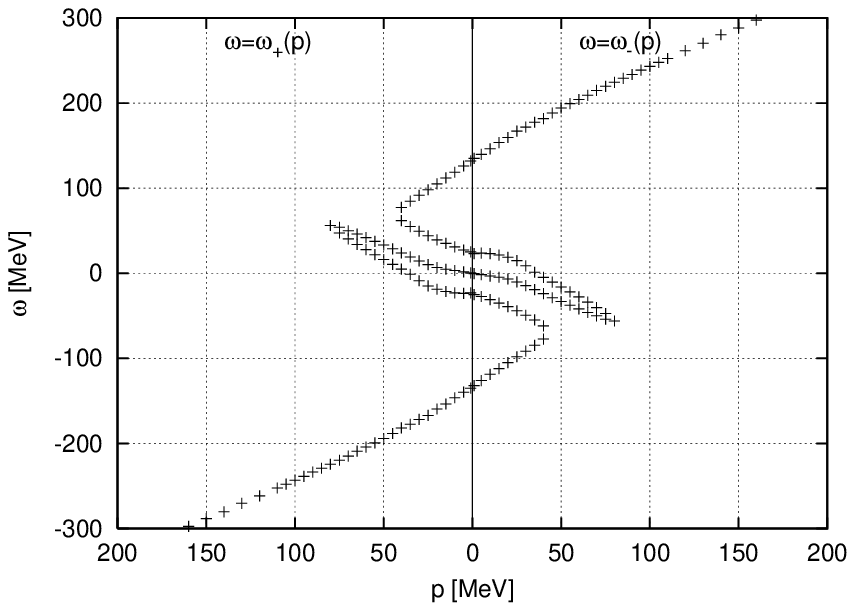, width=190pt}

\epsfig{file=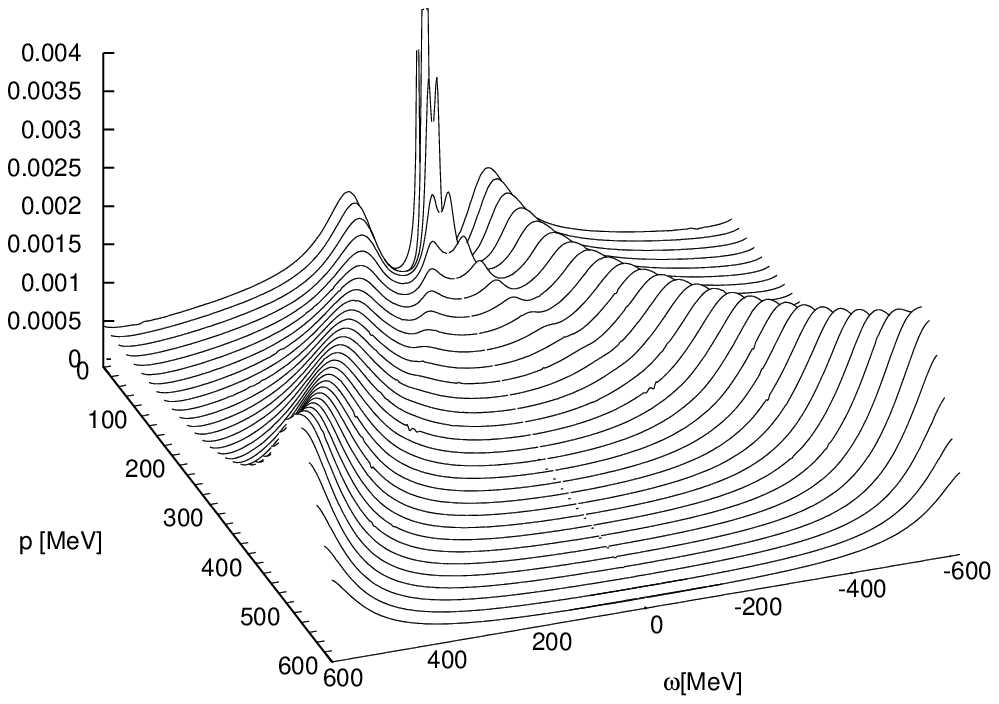, width=190pt}
\epsfig{file=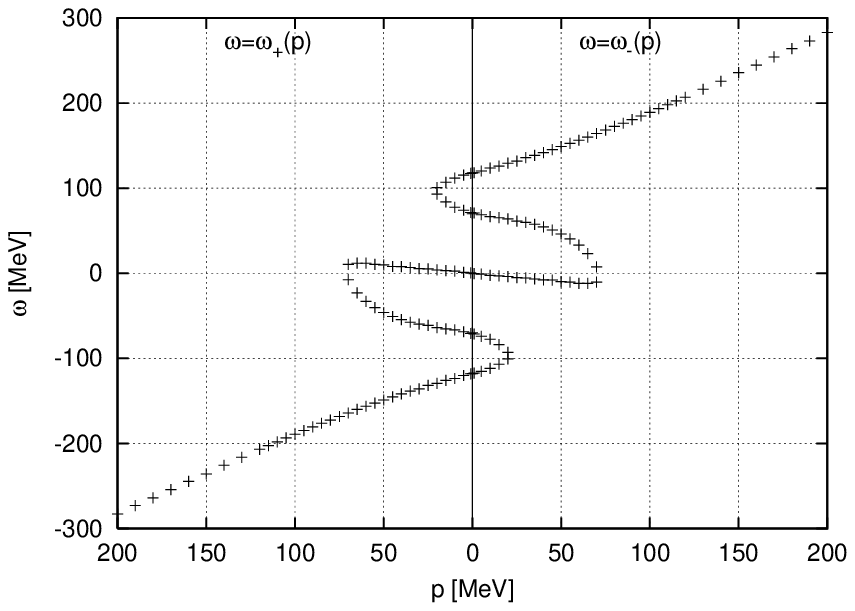, width=190pt}

\epsfig{file=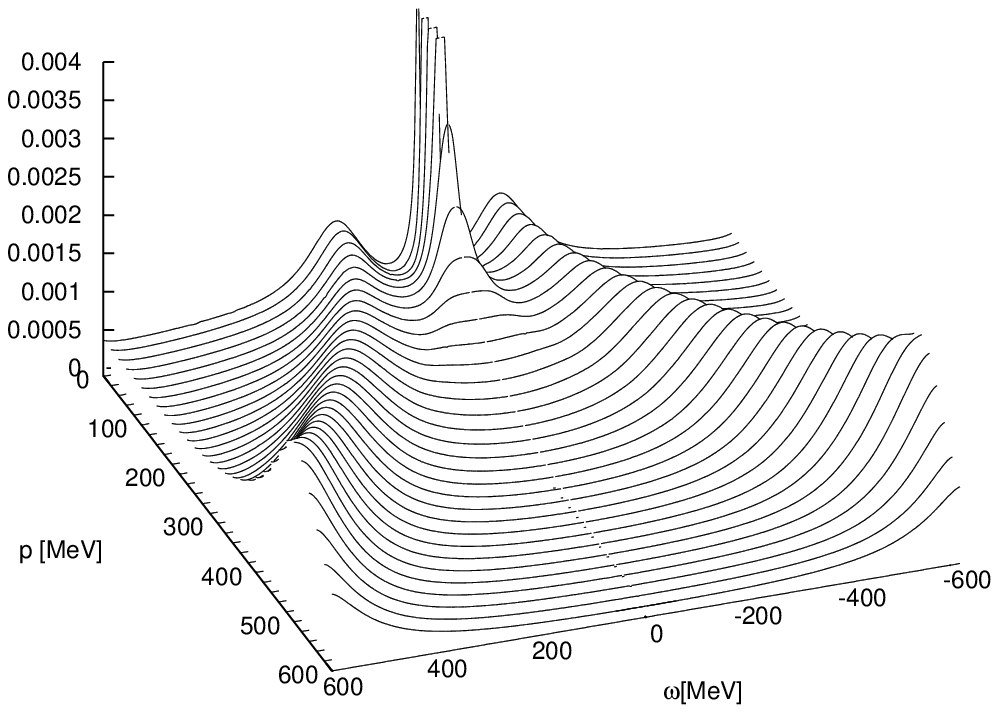, width=190pt}
\epsfig{file=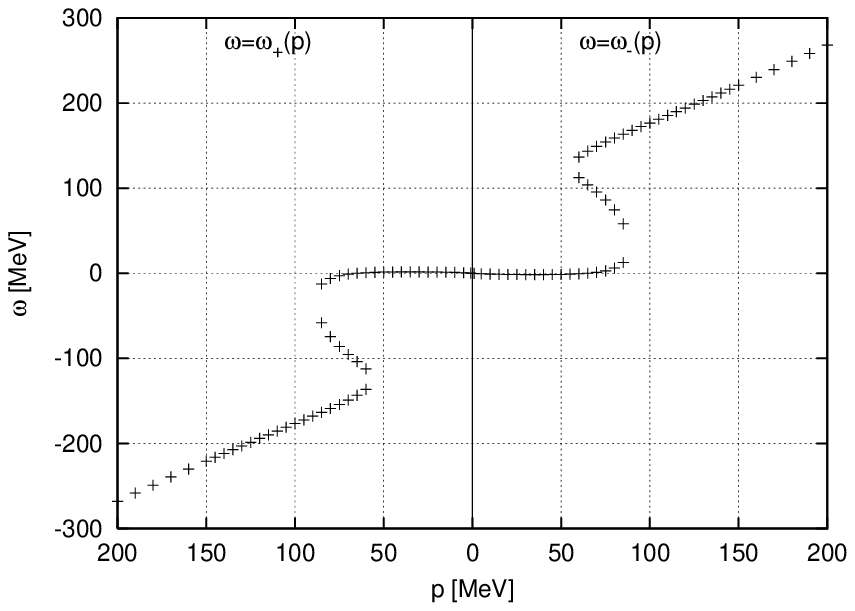, width=190pt}

\epsfig{file=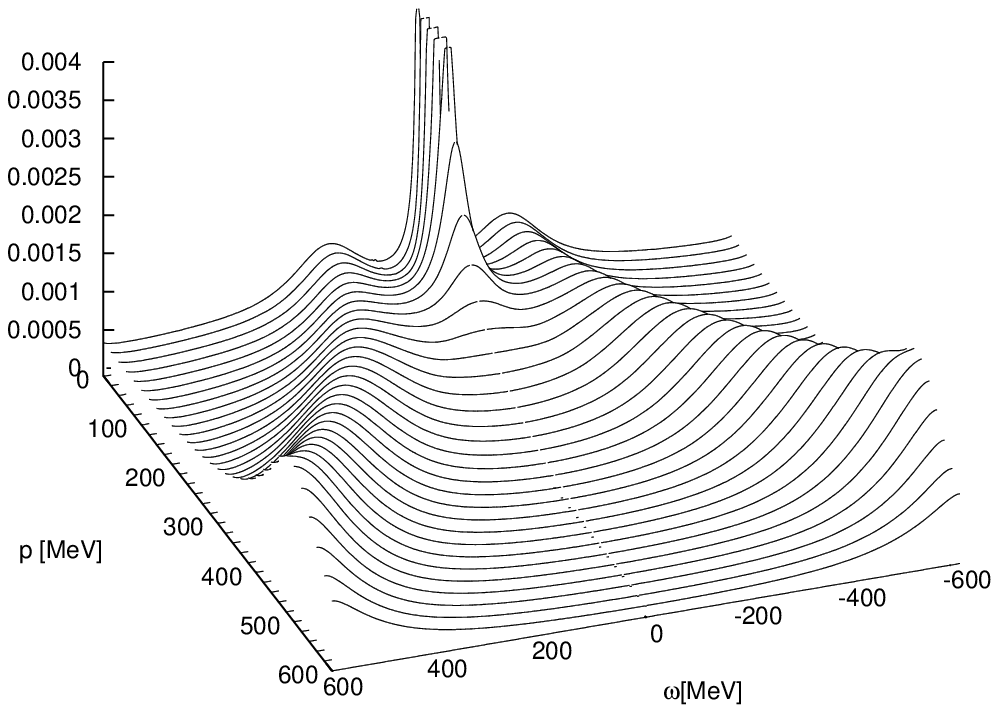, width=190pt}
\epsfig{file=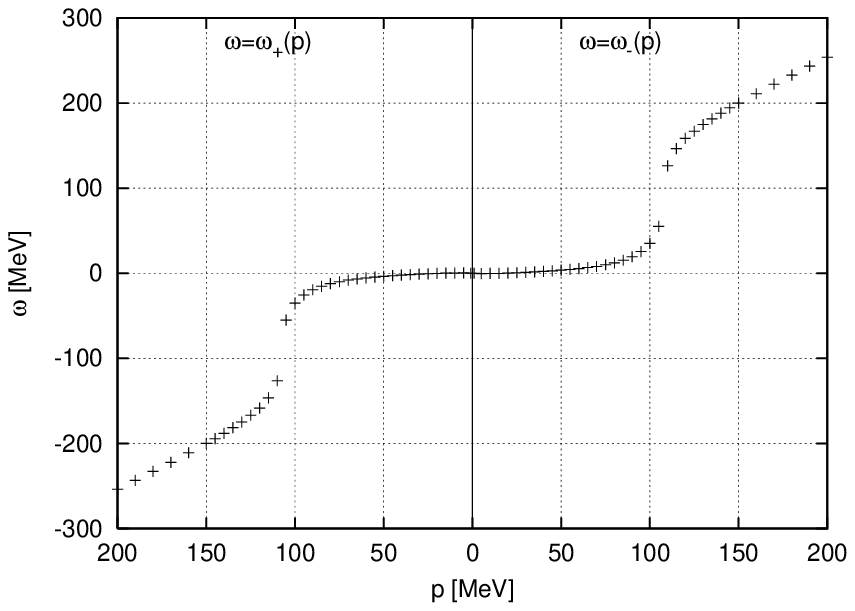, width=190pt}

\caption{The spectral functions $\rho_0(\bm{p},\omega)$ 
and the dispersion relations for
T=196 MeV $(1.01T_C)$,
T=210 MeV $(1.09T_C)$,
T=230 MeV $(1.19T_C)$,
T=250 MeV $(1.29T_C)$ from above.}
\label{spct}
\end{center}
\end{figure}

\begin{figure}[h]
\begin{center}
\epsfig{file=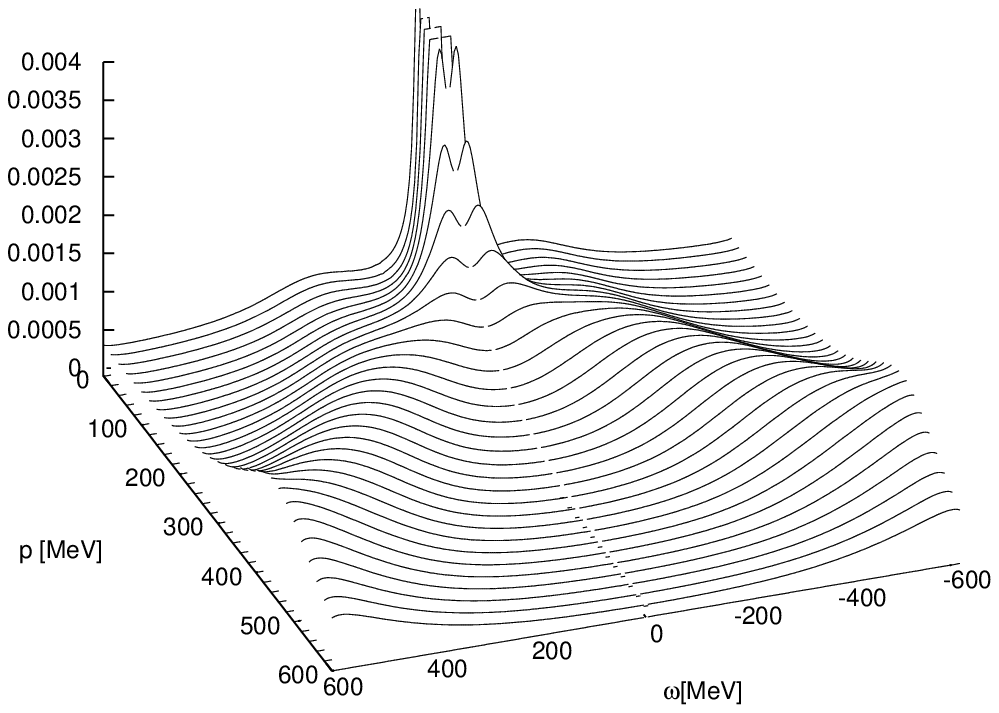, width=190pt}
\epsfig{file=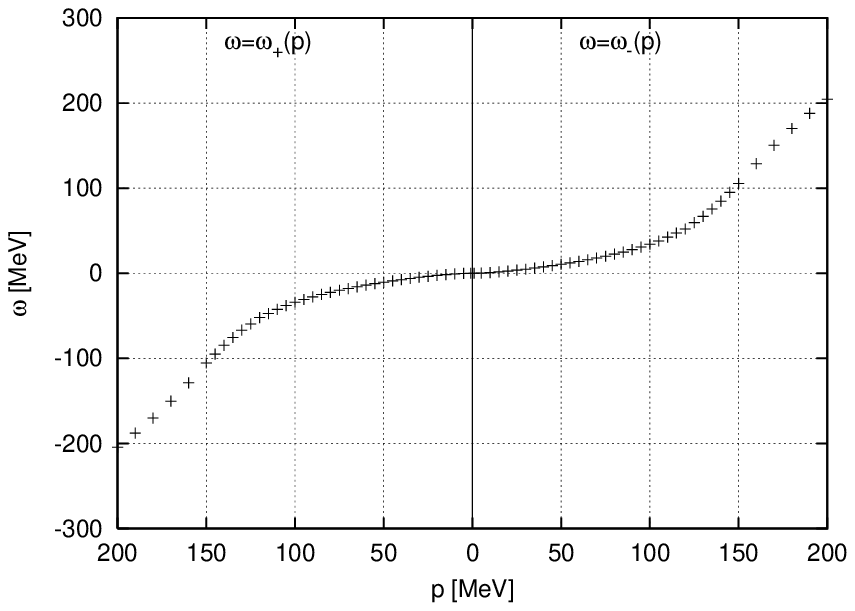, width=190pt}

\epsfig{file=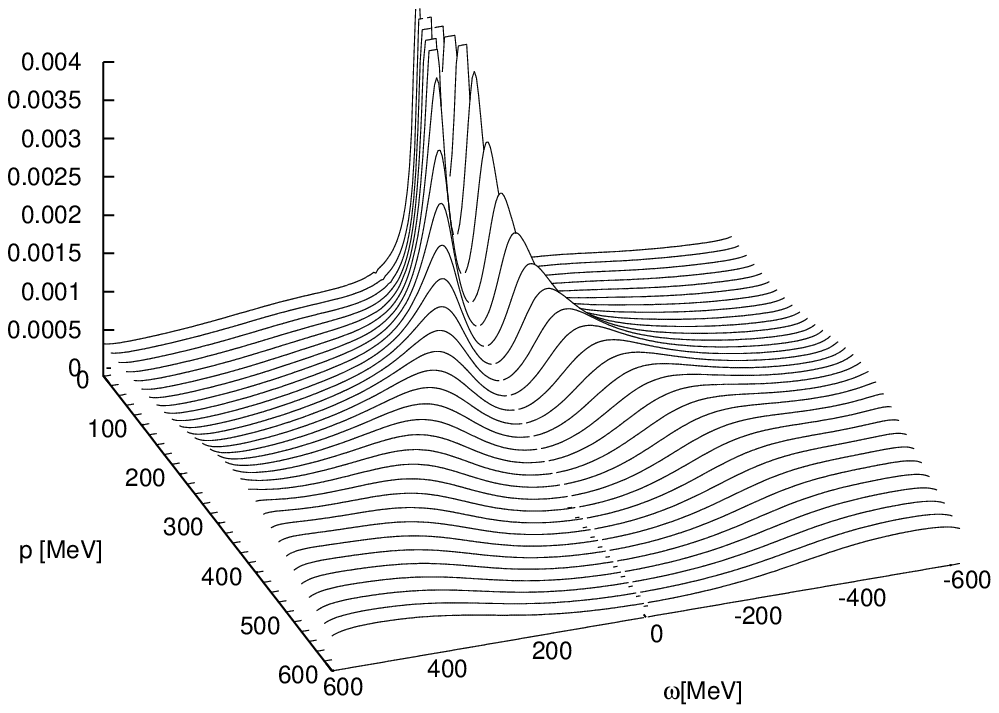, width=190pt}
\epsfig{file=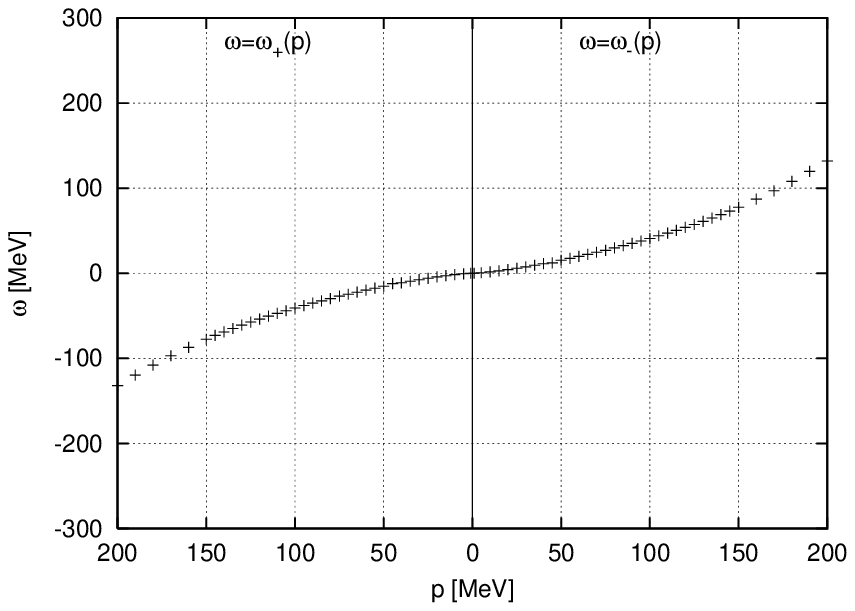, width=190pt}

\caption{The same as Fig.\ref{spct} for 
T=300 MeV $(1.55T_C)$(upper) and
T=400 MeV $(2.07T_C)$(lower).}
\label{disps}
\end{center}
\end{figure}

We show the numerical results of the spectral functions.
The spectral function
$\rho_0(\bm{p},\omega)$ and the corresponding
dispersion relations for each 
temperature and zero chemical potential are shown in 
Figs.\ref{spct} and \ref{disps}.
The dispersion relations, $\omega=\omega_\mp(p)$, are obtained 
from the poles of the retarded Green functions, i.e., the
solutions of the equations
$\omega \mp |\bm{p}| - {\rm Re}\Sigma_\mp + \mu =0$.
For clarity, the dispersion relations for $\omega_-(p)$ and
$\omega_+(p)$ are plotted separately in these figures.
Near $T_C$, one sees several peaks for small $\bm{p}$ and $\omega$.
Although there are ten solutions at $p\approx0$ and $T=196$ MeV
(1.01$T_C$) in the dispersion relation, for example, 
they do not always form peaks in the spectral functions.
This is because the peaks are also formed depending on the
imaginary part of the self-energy.

Now we analyze the peak structure of the spectral functions for
each $T$.
For the high momentum region ($p>100$MeV), 
the number of spectral peaks are two
over all the temperatures.
Their positions are close to those of a free
massless quark and an antiquark, respectively.
Since the fluctuation rapidly reduces at high momentum as shown in
Fig.\ref{soft-tp},
their peaks correspond to nearly free quasi-quark and quasi-antiquark states,
which is also confirmed from the dispersion relations.
The widths of these peaks come from incoherent scatterings off
a quark and an antiquark in the effective 
pair-field propagator $\mathcal{D}$.

At $T\geq 300$ MeV, simple two-peak structures are seen over all
the frequency and the momentum region and their positions are also
close to free quark states.
This is because the fluctuation disappears at these temperatures
and only the incoherent scatterings exist.

\begin{figure}
\begin{center}
\epsfig{file=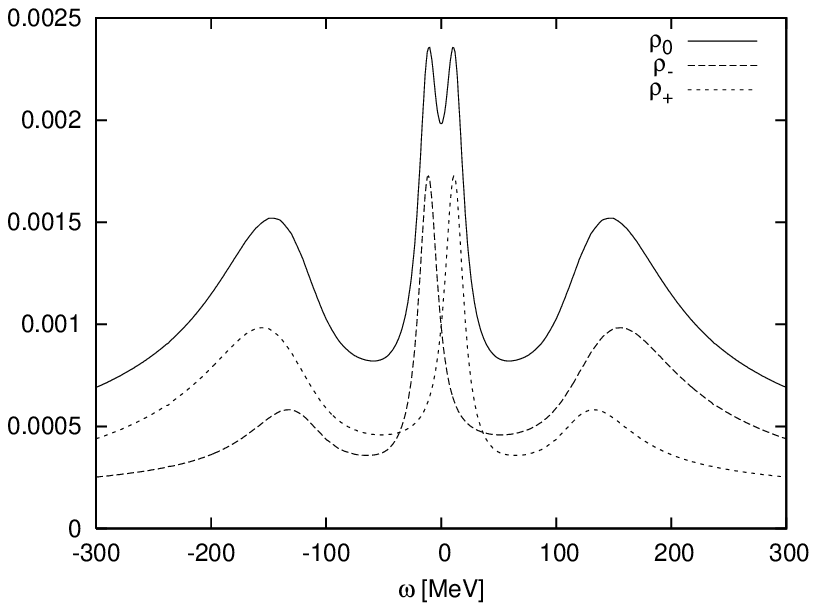, width=200pt}
\caption{The spectral functions at $|\bm{p}|=50$ MeV and
$T=210$ MeV ($1.09T_C$).}
\label{sp210-50}
\end{center}
\end{figure}

\begin{figure}[h]
\begin{center}
\epsfig{file=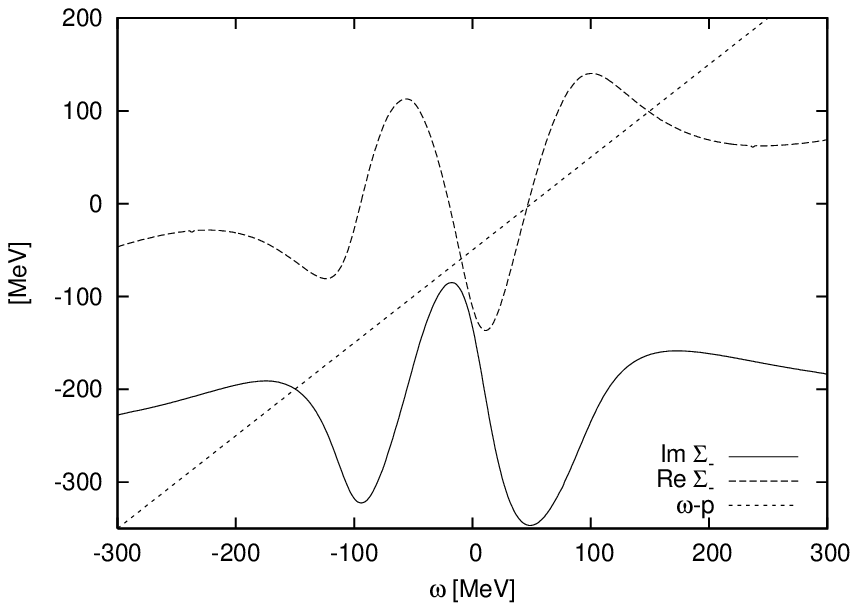, width=200pt}
\caption{The self-energy $\Sigma_-$ at $|\bm{p}|=50$ MeV and
$T=210$ MeV. The line $\omega-|\bm{p}|$ is also shown.}
\label{self210p050}
\end{center}
\end{figure}

\begin{figure}[h]
\begin{center}
\epsfig{file=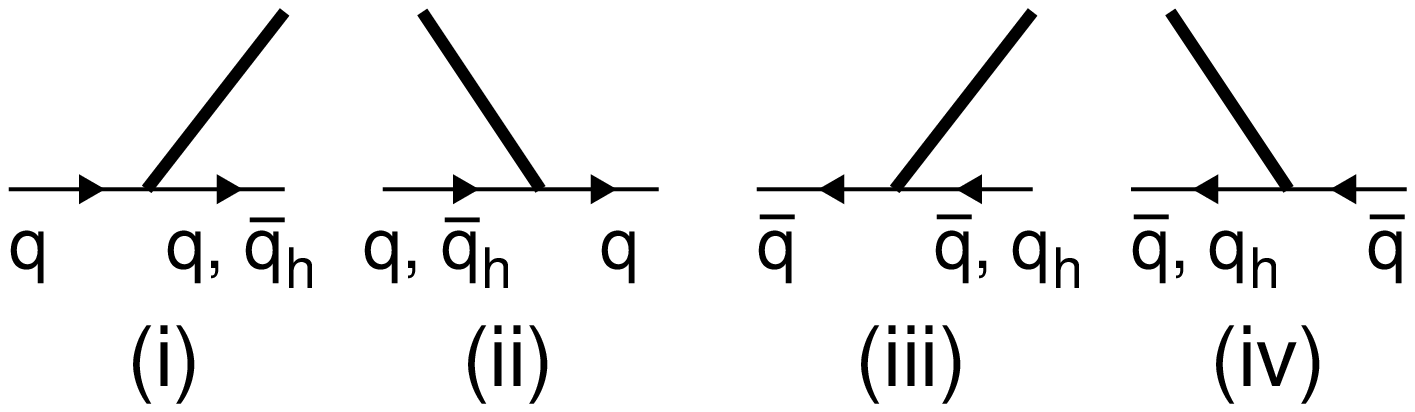, width=200pt}
\caption{Scattering processes of a quark.
$q_h(\bar{q}_h)$ denotes an (anti)quark `hole'.}
\label{diag}
\end{center}
\end{figure}

\begin{figure}[t]
\begin{center}
\epsfig{file=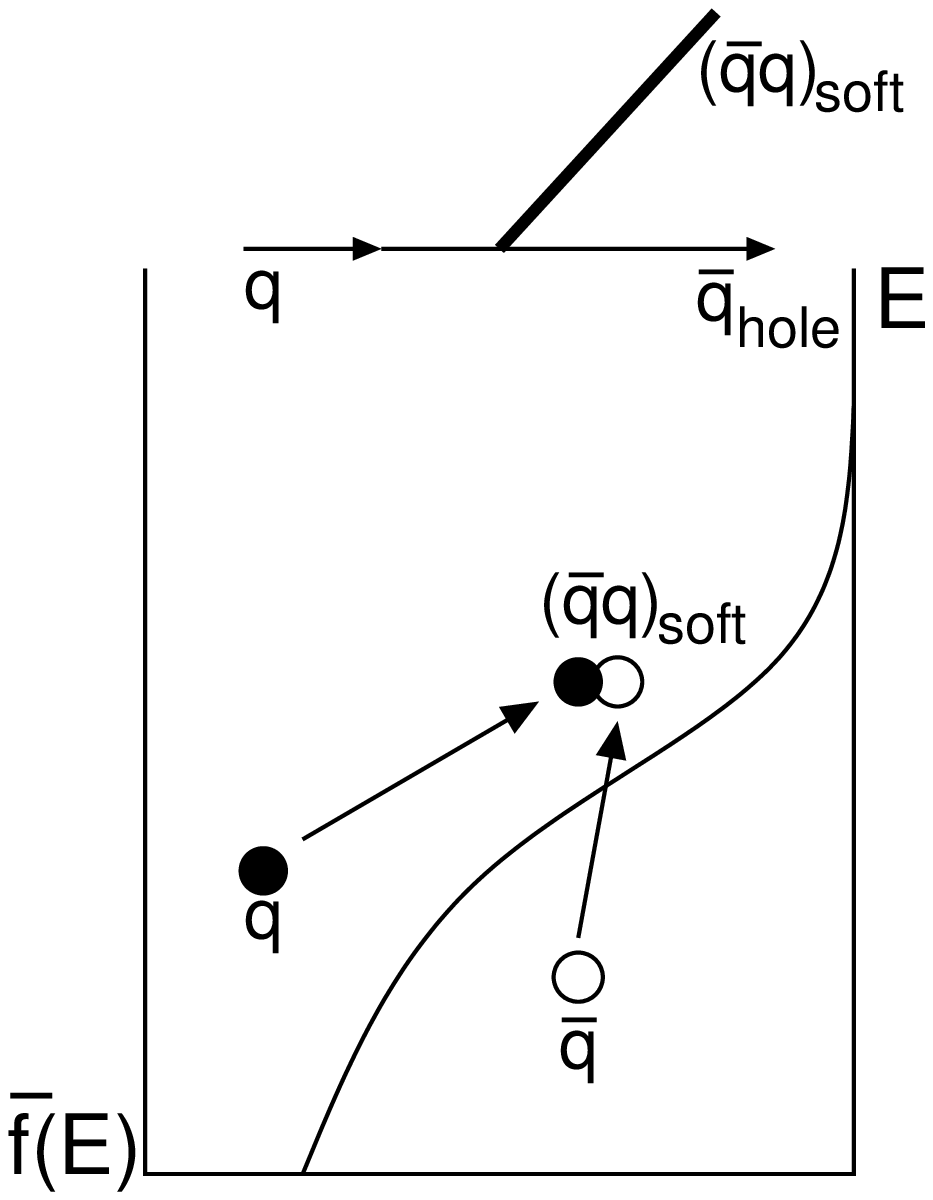, width=100pt}
\epsfig{file=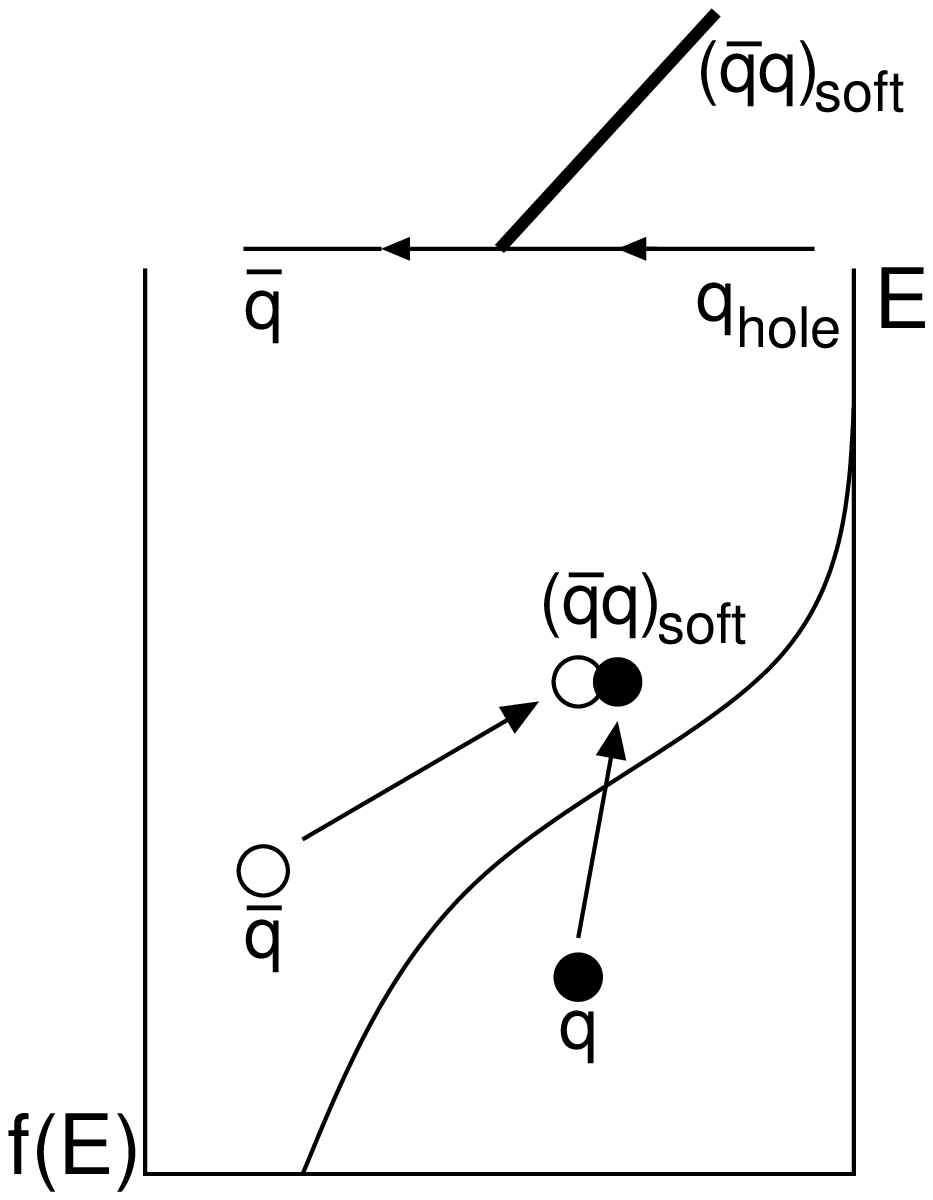, width=100pt}
\caption{Scattering processes of an antiquark(a quark) into
the soft mode and a quark `hole' $q_h$(antiquark `hole' $\bar{q}_h)$.
$f(\bar{f})(E)$ denotes the quark(antiquark) distribution function.
The quark(antiquark) `hole' means the annihilation of a thermally
excited quark(antiquark).}
\label{prc}
\end{center}
\end{figure}

Next we turn to the low frequency and low momentum region near $T_C$
where the several peaks are seen.
As a typical example of such a region, let us consider the
spectral function at $|\bm{p}|=50$ MeV and $T=210$ MeV shown in
Fig.\ref{sp210-50}.
In order to analyze it, it is convenient to
decompose the $\rho_0$ into the $\rho_\mp$ as in this figure.
The $\rho_-$ represents the spectrum for quarks and antiquark
`holes' and the $\rho_+$ for antiquarks and quark `holes'.
Here `holes' mean annihilation of thermally excited particles.
We focus on the $\rho_-$ part and show the corresponding
self-energy $\Sigma_-$ in Fig.\ref{self210p050}.
The $|{\rm Im\ }\Sigma_-|$ has two peaks (large numbers)
around $\omega\approx -100$
and $50$ MeV, where
there exists a strong decay process.
In this case, the peak with $\omega>(<)0$ comes from the scattering
of a quark off the collective (soft) mode with $\omega_s>(<)0$ shown 
in Fig \ref{soft-tp}.
Because the soft mode with $\omega_s>0$ describes the creation process of it,
the peak of the $|{\rm Im\ }\Sigma_-|$ with $\omega>0$ represents the
following processes:
(i) A quark decays into a soft mode and an (on-shell) quark,
$q\to (\bar{q}q)_{\rm soft} + q$, and an antiquark `hole',
$q\to (\bar{q}q)_{\rm soft} + \bar{q}_{\rm h}$.
Similarly, the soft mode with $\omega_s<0$ describes the annihilation process
and thus the peak of the $|{\rm Im\ }\Sigma_-|$ with $\omega<0$ represents
the following processes:
(ii) an (on-shell) quark and a soft mode couple to a quark,
$(\bar{q}q)_{\rm soft} + q\to q$,
and an antiquark `hole' and a soft mode couple to a quark,
$(\bar{q}q)_{\rm soft} + \bar{q}_{\rm h}\to q$.
They are schematically shown in Fig.\ref{diag}.
The origin of the spectral peaks for these processes is
understood from the real
part of the self-energy shown in Fig.\ref{self210p050}.
Recall that the dispersion relation of quarks is obtained from
the solution $\omega-|\bm{p}|-{\rm Re}\Sigma_-(\bm{p},\omega)=0$.
We see that there are three solutions around $\omega\approx -9$ MeV,
$45$ MeV and $148$ MeV from the intersections of Re$\Sigma_-$ and
the line $\omega-|\bm{p}|$.
The solution $\omega\approx -9$ MeV has a sharp peak in the 
spectral function (Fig.\ref{sp210-50}).
The origin of it is a superposition of the scatterings
(i) and (ii) because it lies between two peaks of $|{\rm Im}\Sigma_-|$.
The solution $\omega\approx 45$ MeV has no peak because $|{\rm Im} \Sigma_-|$
is large.
The peak for the solution $\omega\approx 148$ MeV is coming from the
scattering (i). 
There is another peak in $\rho_-$ at $\omega\approx -120$ MeV, which
is not an pole but a remnant of it because for smaller $|\bm{p}|$,
it is an actual pole which comes from the scattering (ii).
The same discussions are possible for the scatterings of an antiquark
from the $\rho_+$ part, whose processes are shown as (iii) and
(iv) in Fig.\ref{diag}.
Therefore, all the peaks at
near $T_C$ can be understood in term of the resonance scatterings
of a quark off a soft mode.
One of the characteristic features of them
is the scatterings into a quark `hole' and an antiquark `hole', 
which appear only at finite temperature and/or density and form
the sharp spectral peaks in the space-like region as in the figure. 
These processes involving 'holes' are schematically shown in 
Fig.\ref{prc}.

For the behavior of the dispersion relations, 
it is instructive to recall the quark spectrum of hot QCD valid
in the high temperature limit\cite{Weldon:1989ys}.
We see that the dispersion relations near $T_C$ and from hot QCD
are similar.
In fact, one can show that our results of the dispersion relations
near $T_C$ are understood as an interpolating behavior between the
free quark dispersion at zero temperature and the dispersion in
the high temperature limit of hot QCD\cite{KKN}.

At finite density, existence of the Fermi surface prevents
antiquarks from exciting thermally and thus the antiquark
`hole' excitation is suppressed.
We have also investigated the quark spectrum near the tricritical 
point and obtained an asymmetric spectrum for the quark and antiquark
sectors. The detail is given in Ref.\cite{KKN2}.

Finally we comment on the related work on the chiral fluctuations.
Recently a possible pseudogap formation in the chiral phase transition
is discussed by several authors\cite{Babaev:2000fj}.
They considered the chiral phase transition due to the phase fluctuation
of the chiral condensate, i.e., existence of the state with 
$\langle \bar{\psi} \psi \rangle=0$ but 
$\langle |\bar{\psi} \psi| \rangle\neq0$ was investigated.
The fluctuations on which we focus in this paper are clearly different
from it: 
We are interested in the states with 
$\langle \bar{\psi} \psi \rangle=\langle |\bar{\psi} \psi| \rangle=0$
but $\langle \bar{\psi} \psi(x)  \bar{\psi} \psi(0) \rangle\neq 0$.
It is also interesting to combine our analysis with the phase
fluctuation and as a result diverse chiral phase structures may appear.

\section{Conclusion} \label{sec:conc}

We have investigated the effects of the fluctuation of the
chiral order parameter, i.e., the soft mode, on the
single-quark spectral 
function near but above the chiral phase transition.
In order to incorporate the strong coupling nature between
quarks, 
we have employed a low energy effective model of QCD,
Nambu--Jona-Lasinio model to evaluate the strength
of the soft mode quantitatively.
In fact, we have shown that the fluctuations of
the chiral condensate make a collective mode
whose spectral peak moves to the origin as 
temperature approaches the critical temperature from above,
as first shown in Ref. \cite{Hatsuda:1984jm}.
Owing to the existence of such fluctuations,
the quark spectrum at low frequency and low momentum
is strongly modified from the free particle one
and shows several peaks.
This peak structure of the spectral function can 
be understood in terms of the resonance scattering of a
quark off a soft mode shown in Fig.\ref{diag}.
The scatterings to a soft mode and an (anti)quark 'hole'
are characteristic of finite temperature
and/or density systems and can form spectral peaks in 
the space-like region.
Similar phenomena occur in hot QCD, where the hard thermal
loop (HTL) approximation can be employed to evaluate the quark
self-energy.
In this case, scattering of a quark of a thermal gluon
leads to plasmino states.
The difference of the dispersion relations between our results and 
the HTL approximated hot QCD is that the former occurs just
above the critical temperature($T_C$) where the fluctuations of
the chiral condensate survive and is insufficient to form
the clear plasmino states because $T_C$ is not so high\cite{KKN}.

The existence of the soft mode is only due to the existence of
the second (or nearly second) order transition and the strength
of it depends on the strong coupling nature of QCD.
Therefore the scattering processes of a quark off the soft mode
must exist universally.
Effects of the finite current quark masses might smear the 
fluctuation, because the order of the transition changes into
crossover at low density.
Furthermore, the fluctuation around the critical end-point
is a different property from that studied in this paper
\cite{Fujii:2003bz}.
Investigation under such a case is remained as a future project.

\end{document}